\documentclass[prd,onecolumn,nofootinbib,superscriptaddress,floatfix,floats,11pt]{revtex4-2} 

\usepackage{graphicx}
\usepackage{amsfonts}
\usepackage{amssymb}
\usepackage{amsbsy}
\usepackage{amsmath}
\usepackage{mathrsfs}
\usepackage{latexsym}
\usepackage{natbib}
\usepackage{bm}
\usepackage[utf8]{inputenc}
\usepackage{subfigure} 
\usepackage{color}
\usepackage[active]{srcltx}
\usepackage{wasysym}
\usepackage{mathbbol}
\usepackage{bigints}
\usepackage{hyperref}
\allowdisplaybreaks
\usepackage{comment}
\usepackage{empheq}
\usepackage{soul}
\usepackage[export]{adjustbox}
\usepackage{gensymb}
\usepackage[utf8]{inputenc}

\newcommand{\bea}[1]{\begin{eqnarray}{#1}}
\newcommand{\eea}{\end{eqnarray}}
\newcommand{\dd}{\mathrm{d}}

\newcommand{\im}{\mathrm{Im}}

\reversemarginpar

\begin{document}

\title{On the Imaginary Part of the Effective Action in de Sitter Spacetime with Different Regularization Schemes}

 \author{Yu Zhou}
 \email{YuZhou\_@buaa.edu.cn}
 \affiliation{Center for Gravitational Physics, Department of Space Science, Beihang University, Beijing 100191, China}
 \author{Hai-Qing Zhang}
 \email{hqzhang@buaa.edu.cn}
 \affiliation{Center for Gravitational Physics, Department of Space Science, Beihang University, Beijing 100191, China}
 \affiliation{Peng Huanwu Collaborative Center for Research and Education, Beihang University, Beijing
100191, China}

\begin{abstract}
\noindent
The imaginary part of the effective action encodes vacuum instability and particle production in the background field. Two standard approaches are commonly used to derive it: the Bogoliubov method and the Green's function method, which are usually expected to agree. However, in de Sitter spacetime they yield different results. We revisit this problem by introducing explicit time and momentum cutoffs in the Green's function representation of the effective action. The apparent discrepancy is found to be due to the different limiting procedures in regularization, which reproduces the Bogoliubov result and the Green's function result respectively. Therefore, the two approaches are understood to be different regularization limits of the same expression, which clarifies the origin of their disagreement.
\end{abstract}

\maketitle
\section{Introduction}
A central insight of quantum field theory is that the vacuum is not empty but exhibits rich physical phenomena, including fluctuations \cite{Casimir:1948dh}, entanglement \cite{Reznik:2002fz}, and, in certain backgrounds, spontaneous particle production \cite{Hawking:1975vcx,Schwinger:1951nm}. Understanding vacuum instability and particle production is essential for quantum field theory in external fields and curved spacetimes, with important applications ranging from the Schwinger effect in strong electromagnetic fields \cite{DUNNE_2005,Gelis_2016} to cosmology in expanding universes \cite{Ford_2021}. A general and unifying framework for describing these effects is provided by the effective action, whose imaginary part quantifies the instability of the vacuum and directly signals particle production \cite{DeWitt:2003pm,Parker_Toms_2009}.

To set the stage, consider the In-vacuum to Out-vacuum scattering amplitude (vacuum persistence amplitude) of a scalar field $\phi $, the effective action is defined by
\begin{equation}\label{defWeff1}
W=-i\log\left \langle \text{Out}  | \text{In}  \right \rangle =-i\log \int \mathcal{D}[\phi]e^{i S[\phi,g]} ,
\end{equation}
where $S[\phi,g]$ is the classical action of the field $\phi$ in a background field $g$ (e.g. gravitational field). The effective action encodes the quantum corrections to the classical dynamics. If the effective action acquires an imaginary part, the persistence probability of the vacuum reads
\begin{equation}
|\langle \text{Out} | \text{In} \rangle|^2 = \exp\left(-2\im W\right),
\end{equation}
which implies a nonzero probability of particle production, $P =1-|\langle \text{Out} | \text{In} \rangle|^2\approx 2\im W $.

Two standard approaches are commonly used to extract the imaginary part of the effective action: the Bogoliubov method and the Green's function method. The Bogoliubov method, based on canonical quantization, expresses the vacuum persistence amplitude in terms of Bogoliubov coefficients that relate the In- and Out-modes of the field. In this framework one finds schematically \cite{Parker_Toms_2009}
\begin{align}\label{defImWB}
\im W_B =\frac{1}{2}\int \dd^{d}\mathbf{x}\int\frac{\dd^d\mathbf{k}}{(2\pi)^d}\log|\alpha_\mathbf{k}|,
\end{align}
where the Bogoliubov coefficient $\alpha_{\mathbf{k}}$ denotes the mode-preserving amplitude relating the In- and Out-modes of momentum $\mathbf{k}$. We consider here a general $(d+1)$-dimensional spacetime, where $d$ denotes the number of spatial dimensions.

In the path-integral based Green's function method, the imaginary part is obtained from the Feynman Green's function in the coincidence limit \cite{Birrell:1982ix} (or equivalently from the heat kernel diagonal \cite{Vassilevich_2003}). A convenient representation is \cite{Akhmedov_2019}
\begin{align}\label{defImWG}
\im W_G=-\frac{1}{2}\int \dd^{d+1} x\sqrt{|g|} \int _{+\infty}^{m^2}\dd \bar{m}^2 \im G_F(x,x;\bar{m}^2),
\end{align}
where $G_F(x,x;\bar{m}^2)$ is the Feynman propagator evaluated at coincident points with a mass parameter $\bar{m}$, and $m$ denotes the physical mass of the field $\phi$. The integration over the auxiliary mass parameter $\bar{m}^2$ then generates the effective action.

In flat spacetime, such as in the Schwinger effect, or in external fields acting for a finite time, the two approaches produce expressions that, while not identical in form, can be reconciled to give the same physical result \cite{Anderson_2018,Akhmedov:2024qvi}. In de Sitter spacetime, however, the situation is more subtle: the two methods lead to apparently different expressions for the imaginary part of the effective action. This discrepancy has led to long-standing discussions about the interpretation of particle production in de Sitter space and the role of vacuum definitions \cite{Gibbons:1977mu,kim2010,Anderson_2014,AKHMEDOV_2010}.

In this work, we revisit this problem by introducing explicit time and momentum cutoffs in the Green’s function representation of the effective action. This unified framework reveals that the two seemingly inconsistent results actually correspond to different limiting procedures in the regularization. One limiting procedure reproduces the Bogoliubov result, while the other yields the usual Green’s function result. In this way, we clarify the origin of the discrepancy and discuss its implications for particle production and effective action techniques in dynamical backgrounds.

\section{Imaginary Part of the Effective Action in de Sitter Spacetime}\label{ImEA}
We consider a free scalar field $\phi$ of mass $m$ in $(d+1)$-dimensional de Sitter spacetime. In the Poincaré patch, the metric is
\begin{align}
\dd s^2 = \frac{1}{H^2 \eta^2}\left(-\dd\eta^2 + \dd\mathbf{x}^2\right),
\end{align}
with conformal time $\eta \in (-\infty,0)$. For simplicity, we set the Hubble parameter $H=1$ in the following. The action of the scalar field is,
\begin{align}
S[\phi] = -\frac{1}{2}\int \dd^{d+1}x \sqrt{|g|}
\left( g^{\mu\nu}\partial_\mu\phi \partial_\nu\phi + m^2 \phi^2 \right).
\end{align}

Quantization proceeds by decomposing the scalar field in Fourier modes, which is possible thanks to the spatial translation invariance in the Poincaré patch
\begin{align}
\phi(\eta,\mathbf{x}) = \int \frac{\dd^d\mathbf{k}}{(2\pi)^d}
\Big[ u_{\mathbf{k}}(\eta)\, e^{i\mathbf{k}\cdot \mathbf{x}}\, a_{\mathbf{k}}
+ u_{\mathbf{k}}^*(\eta)\, e^{-i\mathbf{k}\cdot \mathbf{x}}\, a_{\mathbf{k}}^\dagger \Big],
\end{align}
in which $a_{\mathbf{k}}$ and $a_{\mathbf{k}}^\dagger$ are respectively the annihilation and creation operators, while $u_{\mathbf{k}}(\eta)$ are the time-dependent mode functions. Different choices of mode functions specify different vacua. In particular, the in- and out-modes can be written as \cite{Birrell:1982ix,Akhmedov_2019}
\begin{align}
u_{\text{in},\mathbf{k}}(\eta) &= \frac{\sqrt{\pi}}{2}e^{-\frac{\nu\pi}{2}}(-\eta)^{\frac{d}{2}}H_{i\nu}^{(1)}(-k\eta),\\
u_{\text{out},\mathbf{k}}(\eta) &= \frac{1}{2}\sqrt{\frac{\pi}{2\sinh \pi\nu}}(-\eta)^{\frac{d}{2}}\left[H_{i\nu}^{(1)}(-k\eta)+H_{i\nu}^{(2)}(-k\eta)\right],
\end{align}
where $H^{(1)}_{i\nu}(x)$ and $H^{(2)}_{i\nu}(x)$ denote Hankel functions of the first and second kind, and $k=|\mathbf{k}|$ is the magnitude of the comoving momentum. The parameter $\nu = \sqrt{m^2 - \frac{d^2}{4}}$ is real for $ m>\frac{d}{2}$, which is the case we focus on throughout this work. For related discussions on the light-field case $m < \frac{d}{2}$, see Refs.~\cite{Akhmedov:2024npw,Miller:2025jbz,Jain:2025maa}.
The in- and out-modes, which specify the corresponding vacua in the Poincaré patch, are related by a Bogoliubov transformation,
\begin{align}
u_{\text{out},\mathbf{k}}(\eta) 
= \alpha_{\mathbf{k}}\, u_{\text{in},\mathbf{k}}(\eta)
+ \beta_{\mathbf{k}}\, u_{\text{in},\mathbf{k}}^*(\eta),
\end{align}
where $\alpha_{\mathbf{k}}$ is the mode-preserving amplitude, whose logarithm directly enters the Bogoliubov expression for the imaginary part of the effective action.  The coefficient $\beta_{\mathbf{k}}$ is the mixing amplitude, and its modulus squared $|\beta_{\mathbf{k}}|^2$ encodes the number of particles produced in mode $\mathbf{k}$. The two coefficients are constrained by the normalization condition $|\alpha_{\mathbf{k}}|^2 - |\beta_{\mathbf{k}}|^2 = 1$.  The Bogoliubov coefficients for the in- and out- modes can be evaluated explicitly
\begin{align}
|\alpha_{\mathbf{k}}|^2 =  \frac{1}{1-e^{-2\pi\nu}} ,~~~  |\beta_{\mathbf{k}}|^2 = \frac{1}{e^{2\pi\nu}-1} .
\end{align}
They automatically satisfy the normalization condition and exhibit the thermal Bose–Einstein structure characteristic of particle production in de Sitter spacetime \footnote{Note that this differs from the situation in global spherical coordinates, where the Bogoliubov coefficients of the in- and out-modes exhibit an explicit dependence on the spacetime dimension; in particular, for odd dimensions the in- and out-modes are identical (i.e. $\alpha=1,\beta=0$) \cite{Bousso_2002,Akhmedov_2019}. In this work, we restrict ourselves to the case of Poincaré patch.}. Substituting the explicit expression of $|\alpha_{\mathbf{k}}|^2$ into (\ref{defImWB}), we obtain the Bogoliubov prediction for the imaginary part of the effective action as
\begin{align}\label{ImWB}
\im W_B = -\frac{1}{4}\int \dd^d\mathbf{x}\int \frac{\dd^d\mathbf{k}}{(2\pi)^d}\log\left(1-e^{-2\pi\nu}\right).
\end{align}

In the approach of the Green's function to effective action (\ref{defImWG}), the Feynman propagator $G_F$ is required. In the de Sitter spacetime, the in–out Feynman propagator can be constructed from the mode functions as
\begin{align}\label{MRofGF}
G_F(x,x') = \int \frac{\dd^d\mathbf{k}}{(2\pi)^d}G_{\mathbf{k}}(\eta,\eta')e^{i\mathbf{k}\cdot(\mathbf{x}-\mathbf{x}')},
\end{align}
with
\begin{align}
G_{\mathbf{k}}(\eta,\eta')=\mathcal{N}\left[\theta(\eta-\eta')u_{\text{out},\mathbf{k}}(\eta)u^*_{\text{in},\mathbf{k}}(\eta') +\theta(\eta'-\eta)u_{\text{out},\mathbf{k}}(\eta')u^*_{\text{in},\mathbf{k}}(\eta)\right],
\end{align}
where $\theta(x)$ denotes the Heaviside step function and $\mathcal{N}=\sqrt{1-e^{-2\pi\nu}}$ is the normalization factor. Performing the Fourier integration, one finds that the propagator depends only on the de Sitter–invariant hyperbolic distance $z=\frac{\eta^2+\eta'^2-(\mathbf{x}-\mathbf{x'})^2}{2\eta\eta'}-i\epsilon$ (where $\epsilon>0$ is the infinitesimal parameter in the Feynman $i\epsilon$ prescription fixing the analytic continuation of the propagator),  as expected from the maximal symmetry of the de Sitter spacetime. The result can be expressed in closed form as \cite{Akhmedov_2019,Fukuma_2013}
\begin{align}
G_F(z)=\frac{e^{-i\pi(d-1)}}{(2\pi)^{\frac{d+1}{2}}}(z^2-1)^{-\frac{d-1}{4}}Q_{-\frac{1}{2}+i\nu}^{\frac{d-1}{2}}(z),
\end{align}
where $Q^\mu_\nu(x)$ is the associated Legendre function of the second kind. In the coincidence limit  $z=1$, the in–out Feynman propagator diverges and must be regularized. Following the previous work \cite{Akhmedov_2019}, dimensional regularization together with the modified minimal subtraction ($\overline{\text{MS}}$) scheme can be employed to remove the divergence. After renormalization, the finite part of the coincident propagator exhibits a characteristic dependence on the space–time dimensions: it develops a nonvanishing imaginary contribution in even dimensions, while the imaginary part vanishes in odd dimensions \cite{Akhmedov_2019,zhou2024}, i.e., 
\begin{align}
\im G^{\text{ren}}_F(x,x) &=-\frac{(-1)^{\frac{d+1}{2}}}{(4\pi)^{\frac{d+1}{2}}\Gamma\left( \frac{d+1}{2} \right)}e^{-\pi\nu}\Big|\Gamma\left(\frac{d}{2} +i\nu\right)\Big|^2  \ , \ &  \text{for even $d+1$ dimensions}, \\
\im G^{\text{ren}}_F(x,x) &= 0  \ , \ & \text{for odd $d+1$ dimensions},
\end{align}
where $\Gamma(x)$ denotes the Gamma function. Thus, according to (\ref{defImWG}), a nonvanishing imaginary part of the effective action arises only in even space–time dimensions, and is given by
\begin{align}\label{ImWG}
\im W_G=\frac{(-1)^{\frac{d+1}{2}}}{(4\pi)^{\frac{d+1}{2}}\Gamma\left( \frac{d+1}{2} \right)}\int\dd^{d+1}x\sqrt{|g|}\int_{+\infty}^{\nu}\dd \mu\mu e^{-\pi\mu}\Big|\Gamma\left(\frac{d}{2} +i\mu\right)\Big|^2 \ , \ (d+1=2,4,6,\cdots) .
\end{align}
In the above equation, we have changed the integration variable from $\bar{m}^2$ to $\mu = \sqrt{\bar{m}^2-\frac{d^2}{4}}$, with $\dd \bar{m}^2 = 2\mu\dd \mu$. The resulting integration over $\mu$ can be evaluated exactly \cite{zhou2024}.

This feature leads to a fundamental discrepancy between the Green’s function approach and the Bogoliubov approach about particle production: the existence of an imaginary part, and hence of particle production, becomes ambiguous depending on the method employed. Moreover, even in even space–time dimensions where the Green’s function approach produces a nonvanishing imaginary part, the sign of the imaginary part alternates with dimensions. As a result, the imaginary part of the effective action changes sign with dimensions. When it turns negative, the persistence probability blows up, signaling a puzzling loss of unitarity.
These striking discrepancies have led to onging discussions in the literatures, with various attempts to reconcile the two approaches \cite{Mottola:1984ar,Anderson_2014,Anderson_2018,Kim:2011rx,Akhmedov:2024qvi}. To the best of our knowledge, however, no unified resolution has yet been established in which the complete expression obtained by one method is derived from the other. We observe that the integral representations underlying the two results involve fundamentally different integration variables, while they share an important structural feature: in both cases, the integrand itself is independent of the integration variable, leading to divergent expressions. These observations provide a hint towards unification: the two approaches may be reconciled within a common integral representation that incorporates all relevant variables, differing only by the regularization prescriptions applied to the resulting divergent integrals.

The Green’s function representation of the effective action provides precisely a unifying framework. By substituting (\ref{MRofGF}) into (\ref{defImWG}), we obtain an expression that incorporates all integration variables in a single formula
\begin{align}\label{Weff}
W=-\frac{1}{2}\int \dd^{d} \mathbf{x}\int_{-\infty}^{0} d\eta \sqrt{|g(\eta)|} \int _{+\infty}^{m^2}\dd \bar{m}^2 \int_{-\infty}^{+\infty}\frac{\dd^d\mathbf{k}}{(2\pi)^d} G_{\mathbf{k}}(\eta,\eta),
\end{align}
with $\sqrt{|g(\eta)|}=\frac{1}{(-\eta)^{d+1}}$ and 
\begin{align}
G_{\mathbf{k}}(\eta,\eta)=\frac{1}{2} \pi  (-\eta )^d J_{i \nu }(k \eta ) H_{-i \nu }^{(2)}(-k \eta ).
\end{align}
Here we set $\theta(0)=\frac{1}{2}$ and $J_{i\nu}(x)$ denotes the Bessel function of the first kind. Due to the spatial translation invariance, the integrand is independent of the spatial coordinates, so the spatial integral simply contributes an overall volume factor $V_d=\int \dd^d \mathbf{x}$. The remaining integrations over conformal time and comoving momentum are divergent. In order to obtain a finite result, we can introduce two explicit cutoffs: a late-time cutoff $\tau \to 0^{-}$ in the time integration and a momentum cutoff $\Lambda \to +\infty$ in the comoving momentum integration, respectively. Since the expression $G_{\mathbf{k}}(\eta,\eta)$ depends only on the magnitude of the comoving momentum, it is convenient to rewrite the momentum integration in spherical coordinates. The effective action can then be re-written as
\begin{align}\label{WeffdS}
 W=\frac{V_d}{2}\frac{ 2\pi ^{\frac{d}{2}} }{(2 \pi )^d \Gamma \left(\frac{d}{2}\right)}\int _{\nu}^{+\infty}\dd \mu  \int _{-\infty}^{\tau}\dd\eta\int_0^{\Lambda}\dd k\frac{\pi\mu k^{d-1}}{(-\eta)}  J_{i \mu }(k \eta ) H_{-i \mu }^{(2)}(-k \eta ).
\end{align}
As before, we change the integration variable $\bar{m}^2$ to $\mu$. In this regulated integration, the entire imaginary contribution originates from the factor $J_{i \mu }(k \eta ) H_{-i \mu }^{(2)}(-k \eta )$. Making use of the following identity
\begin{align}
\im\left[ J_{i \mu }(k \eta ) H_{-i \mu }^{(2)}(-k \eta )\right] = \frac{i}{2}  \left(\coth (\pi  \mu )-1\right) \left(J_{i \mu }(-k \eta )^2-J_{-i \mu }(-k \eta )^2\right),
\end{align}
the imaginary part of the effective action can be written as
\begin{align}\label{ImWint}
\im W=i\frac{ V_d}{4} \frac{ 2\pi ^{\frac{d}{2}} }{(2 \pi )^d \Gamma \left(\frac{d}{2}\right)}\int _{\nu}^{+\infty}\dd \mu \int _{-\infty}^{\tau}\dd\eta\int_0^{\Lambda}\dd k\frac{\pi\mu\left(\coth (\pi  \mu )-1\right) k^{d-1}}{(-\eta)}  \left(J_{i \mu }(-k \eta )^2-J_{-i \mu }(-k \eta )^2\right).
\end{align}
The integration over $\eta$ and $k$ can be evaluated to a closed form in terms of the generalized hypergeometric function ${}_2F_3$.  The explicit result is given below, while the detailed steps of the derivation are presented in Appendix~\ref{AppendixA},
\begin{align}\label{ImWdS}
\im W=\frac{V_d}{4} \frac{2\pi^{d/2}}{(2\pi)^d\Gamma(\frac{d}{2})}\frac{\Lambda^d}{d}
\int _{\nu}^{+\infty}\dd \mu \pi\mu\left(\coth (\pi  \mu )-1\right)\left[\frac{1}{\mu}+\mathcal{F}(\mu,d,\lambda)+\tilde{\mathcal{F}}(\mu,d,\lambda)\right],
\end{align}
where $\mathcal{F}(\mu,d,\lambda)$ is the function arising from the hypergeometric sector ( $\tilde{\mathcal{F}}(\mu,d,\lambda)$ is the complex conjugate of $\mathcal{F}(\mu,d,\lambda)$ )
\begin{align}\label{FforGHFs}
\mathcal{F}(\mu,d,\lambda)=\frac{(\lambda/2)^{-2i\mu}}{\Gamma(1-i\mu)^2} \left(\frac{\mathcal{F}_1(\mu,d,\lambda)}{2\mu+id} -\frac{\mathcal{F}_2(\mu,\lambda)}{2\mu}\right),
\end{align}
with $\mathcal{F}_1(\mu,d,\lambda)$ and $\mathcal{F}_2(\mu,d,\lambda)$ defined in terms of the generalized hypergeometric functions
\begin{align}
\mathcal{F}_1(\mu,d,\lambda)&=\, _2F_3\left(\frac{1}{2}-i \mu ,\frac{d}{2}-i \mu ;1-i \mu ,\frac{d}{2}-i \mu +1,1-2 i \mu ;-\lambda ^2\right), \label{GHF1} \\
\mathcal{F}_2(\mu,\lambda)&=\, _2F_3\left(\frac{1}{2}-i \mu ,-i \mu ;1-i \mu ,1-i \mu ,1-2 i \mu ;-\lambda ^2\right). \label{GHF2}
\end{align}
Here we have defined the dimensionless parameter,
\begin{align}
\lambda \equiv -\tau\Lambda \to 0^+ \times \infty^+ \in \mathbb{R}^+ ,
\end{align}
which will be referred to as the trade-off parameter (or simply, the trade-off). It will play a central role in the following evaluations, characterizing the interplay between the time cutoff $\tau$ and the momentum cutoff $\Lambda$. Since it formally appears as a $0\times \infty$ indeterminate product, $\lambda$ can in principle take any positive real value, depending on the renormalization prescription. In this sense, the trade-off $\lambda$ parametrizes different limiting procedures in the regularization, and its dependence encapsulates the essential ambiguity underlying the two approaches.

In the following, we will complete the final integration in (\ref{ImWdS}) by taking different limiting values of $\lambda$, to explicitly demonstrate how one recovers the imaginary part of the effective action as obtained from the Bogoliubov method and from the Green’s function method, respectively.

$\bullet$ \underline{$\lambda\to 0^{+}$} : Bogoliubov Result

We first examine the case with the limit $\lambda \to 0^{+}$. It can be regarded as the situation where, in the course of evaluating (\ref{Weff}), the comoving momentum remains finite while the late–time cutoff $\tau \to 0$ dominates the behavior of the integration. In practice, this corresponds to performing the $\eta$–integration first and neglecting terms that oscillate rapidly with $k\tau$ as well as higher–order contributions, retaining only the constant part. In this limit, the generalized hypergeometric functions simplify to $\mathcal{F}_1(\mu,d,\lambda\to 0)=\mathcal{F}_2(\mu,d,\lambda\to 0)=1$. Consequently, the equation (\ref{ImWdS}) reduces to
\begin{align}
\im W=\frac{V_d}{4} \frac{2\pi^{d/2}}{(2\pi)^d\Gamma(\frac{d}{2})}\frac{\Lambda^d}{d}
\int _{\nu}^{+\infty}\dd \mu \left[\pi\left(\coth (\pi  \mu )-1\right)-\frac{de^{-\pi\mu}}{\mu\sqrt{4\mu^2+d^2}}\cos\left(-2\mu\log(\lambda/2)+\varphi \right)\right],
\end{align}
where the regular phase $\varphi(\mu,d)=\arctan\left( \frac{2\mu}{d}\right)-2\arg\Gamma(1-i\mu)$ does not depend on $\lambda$. In the strict $\lambda \to 0^{+}$ limit, the second term in the integrand oscillates rapidly in $\mu$ with frequency $-2\log(\lambda/2)\to +\infty$, while its amplitude is exponentially suppressed by $e^{-\pi\mu}$. By the Riemann–Lebesgue lemma \cite{Haber2018RiemannLebesgue}, this term gives a vanishing contribution upon the integration over $\mu$. The expression therefore simplifies to
\begin{align}
\im W=\frac{V_d}{4} \frac{2\pi^{d/2}}{(2\pi)^d\Gamma(\frac{d}{2})}\frac{\Lambda^d}{d}
\int _{\nu}^{+\infty}\dd \mu \pi\left(\coth (\pi  \mu )-1\right).
\end{align}
The remaining integration can be readily evaluated, yielding
\begin{align}\label{RecoverB}
\im W(\lambda\to 0^+)=-\frac{V_d}{4} \frac{2\pi^{d/2}}{(2\pi)^d\Gamma(\frac{d}{2})}\frac{\Lambda^d}{d}\log\left(1-e^{-2\pi\nu} \right) =\im W_B.
\end{align}
The prefactor $\frac{2\pi^{d/2}}{(2\pi)^d\Gamma(\frac{d}{2})}\frac{\Lambda^d}{d}$ corresponds to the momentum integration $\int_{k\le\Lambda}\frac{\dd^d\mathbf{k}}{(2\pi)^d}$, therefore, it correctly reproduces the results obtained via the Bogoliubov method (\ref{ImWB}).

$\bullet$ \underline{$\lambda\to +\infty$} : Green's function Result

We now turn to the case with the opposite limit $\lambda \to +\infty$. In this situation, the momentum cutoff $\Lambda \to +\infty$ dominates the behavior of the integration (\ref{Weff}). In practice, this corresponds to performing the $k$–integration first, thereby obtaining the coincident limit of the propagator before carrying out the remaining integrations, as in (\ref{defImWG}). In this regime, the generalized hypergeometric functions \eqref{GHF1} and \eqref{GHF2} exhibit non–trivial asymptotic behavior for large $\lambda$, and the last two terms in the integration (\ref{ImWdS}) thus give rise to
\begin{align}\label{InfLimofF}
&\frac{1}{\mu}+\mathcal{F}(\mu,d,\lambda\to+\infty)+\tilde{\mathcal{F}}(\mu,d,\lambda\to+\infty)  \notag\\
=& \lambda ^{-d}\frac{\sinh(\pi\mu)}{\sqrt{\pi}\Gamma(\frac{d+1}{2})\Gamma(1-\frac{d}{2})}\Big|\Gamma\left(\frac{d}{2} +i\mu\right)\Big|^2 +\sum_{n=1}^{d}\lambda^{-n}\left[a_n(\mu,d)\cos(2\lambda)+ b_n(\mu,d)\sin(2\lambda) \right]\\
&+\mathcal{O}(\lambda^{-(d+1)}) .\notag
\end{align}
The coefficients $a_n(\mu,d)$ and $b_n(\mu,d)$ originate from the large-argument expansion of the generalized hypergeometric functions ${}_2F_3$ in (\ref{FforGHFs}), which can in principle be constructed systematically according to Ref.\cite{DLMF}. These coefficients produce oscillatory contributions of the form $\lambda^{-n}\cos(2\lambda)$ and $\lambda^{-n}\sin(2\lambda)$. Their appearance reflects the sensitivity of the regulated integral to the hard cutoffs. As reported in Ref.\cite{AAD_2025}, such sharp switchings generally introduce interference terms tied to the abrupt nature of the cutoff, rather than to the intrinsic dynamics of particle production. In the present calculation, the oscillatory structures exactly represent the cutoff-induced interference between the late-time cutoff at $\tau$ and the momentum cutoff at $\Lambda$. Although they can dominate the integrand’s behavior for large but finite $\lambda$, they do not yield a smooth limit as $\lambda\to+\infty$. From a physical perspective, we retain only the smoothly varying component, represented by the non–oscillatory term proportional to $\lambda^{-d}$. Substituting this part into the integration (\ref{ImWdS}), we get
\begin{align}\label{RecoverG}
\im W(\lambda\to+\infty)= \frac{1}{2\pi^{d/2}\Gamma(-\frac{d}{2})\Gamma(d+1)}V_d\frac{(-\tau)^{-d}}{d}\int_{+\infty}^{\nu} \dd\mu \mu e^{-\pi\mu}\Big|\Gamma\left(\frac{d}{2} +i\mu\right)\Big|^2.
\end{align}
Here we have reinstated the explicit dependence on the time cutoff $\tau$ through the relation $\lambda=-\tau\Lambda$, therefore, the factor $\frac{(-\tau)^{-d}}{d}$ corresponds to the time integration $\int_{-\infty}^{\tau} \dd \eta \sqrt{|g(\eta)|}$. Together with the spatial volume factor $V_d$, they combine into the full spacetime volume integral $\int \dd^{d+1}x \sqrt{|g|}$. It is straightforward to see that in odd space–time dimensions ($d+1$ odd, i.e. $d$ even), the prefactor vanishes since $\tfrac{1}{\Gamma(-d/2)}=0$, implying that the imaginary part of the effective action disappears in this case. In contrast, in even space–time dimensions the prefactor takes the form
\begin{align}
\frac{1}{2\pi^{d/2}\Gamma(-\frac{d}{2})\Gamma(d+1)}
= \frac{(-1)^{\tfrac{d+1}{2}}}{(4\pi)^{\frac{d+1}{2}}\Gamma\left(\frac{d+1}{2}\right)} \ , \ (d+1=2,4,6\cdots),
\end{align}
which coincides precisely with the coefficient in (\ref{ImWG}). Therefore, in the limit $\lambda \to +\infty$ we  obtain
\begin{align}
\im W(\lambda \to +\infty) = \im W_G ,
\end{align}
which correctly reproduces the imaginary part of the effective action derived from the Green’s function method (\ref{ImWG}), up to rapidly oscillating cutoff–induced terms that should not exist in the physical $\lambda\to+\infty$ limit.

\section{Conclusions and Discussions}\label{conclusion}

In this work, we revisited the long-standing discrepancy between the Bogoliubov method and the Green’s function method for computing the imaginary part of the effective action in de Sitter spacetime. While both approaches are well established and generally consistent in backgrounds that admit asymptotic Minkowskian regions, or in external fields acting for a finite time, their results in de Sitter space differ dramatically. The Bogoliubov method predicts a non–vanishing and positive imaginary part corresponding to thermal particle production, whereas the Green’s function method yields vanishing results in odd spacetime dimensions and results alternate in sign in even dimensions. 

In this paper, by introducing explicit cutoffs in both conformal time and comoving momentum, we provided a unified representation of the effective action (\ref{WeffdS}). In this regulated framework, a single trade-off parameter $\lambda = -\tau \Lambda$ controls the interplay between the late–time cutoff $\tau \to 0^-$ and the momentum cutoff $\Lambda \to +\infty$. Different limiting values of $\lambda$ correspond to distinct regularization prescriptions:
\begin{itemize}
\item In the limit $\lambda \to 0^+$, the Bogoliubov result is exactly reproduced (\ref{RecoverB});
\item In the opposite limit $\lambda \to +\infty$, we recover the Green’s function result, including its characteristic dependence on spacetime dimension (\ref{RecoverG}).
\end{itemize}

This observation clarifies that the two seemingly conflicting answers are not mutually inconsistent, but rather they arise from different regularization limits of the same underlying expression. In other words, the discrepancy originates from distinct computational procedures:
\begin{itemize}
\item In the Bogoliubov approach, one implicitly smooths out the time evolution. In practice, this is reflected in first performing the time integration in (\ref{WeffdS}), and treating the removal of oscillatory contributions as a way of renormalizing the resulting expression;

\item In the Green’s function approach, one instead makes direct use of the coincidence limit of the Feynman propagator. This amounts to first carrying out the momentum integration in (\ref{WeffdS}), and then renormalizing the outcome through an appropriate regularization scheme such as dimensional regularization, in which the regularization procedure automatically suppresses the oscillatory boundary interference effects appearing in (\ref{InfLimofF}).
\end{itemize}

Both methods yield finite answers, but at the cost of suppressing certain potential contributions, which explains why they lead to different results in de Sitter spacetime.

Our analysis should be compared with that of Akhmedov et al. \cite{Akhmedov:2024qvi}, who attributed the discrepancy to the additional contributions arising from the vacuum wave functionals in the functional integral (\ref{defWeff1}), beyond merely specifying the in–out boundary conditions. In our cutoff framework, such contributions are not taken into account explicitly, however, we get consistent results in opposite limits. It seems that including the vacuum wave functionals are not very indispensable, however, on the other hand it will be interesting in future to examine whether including them can lead to a unique, ambiguity-free result. Our approach also differs from the trick used in earlier works \cite{Mottola:1984ar,Anderson_2014,Anderson_2018}, where a divergent momentum integral was effectively reinterpreted as a divergent time integral through scaling relations connected by physical cutoffs (see \cite{Akhmedov:2024qvi} for examples). While that procedure provides a way to handle divergences, it does not precisely reproduce both outcomes. By contrast, our framework unifies the two limiting results within a single regulated expression, rather than merely substituting one source of divergence for another.

The trade-off parameter $\lambda$ can be interpreted as the maximum physical momentum that remains accessible at late times in de Sitter spacetime (with Hubble scale $H=1$). From the structure of the integral (\ref{Iint} in Appendix \ref{AppendixA}), the modes whose physical momenta lie below this late-time accessible scale $\lambda$ contribute to the effective action in the form as in the Green’s function approach, while the modes with physical momenta above $\lambda$ enter through the Bogoliubov-type contribution. These two classes of modes contribute to the imaginary part of the effective action through distinct integral representations, which explains why the two calculation schemes generally yield different results. If one assumes that cosmological redshift drives all physical momenta to zero at late times, the effective action reduces to the Bogoliubov result. Conversely, if one assumes that a late-time observer still has access to the full range of physical momenta, the Green's function result will be obtained. Which prescription is physically appropriate requires careful consideration of the specific observational and theoretical circumstances. Clarifying this distinction is an interesting direction for future work.

From a physical viewpoint, the Bogoliubov prescription has the advantage of yielding a manifestly positive imaginary part, naturally interpreted as particle production probabilities, whereas the Green’s function prescription shows a more intricate dependence on dimension and may even produce unphysical results such as negative probabilities. This suggests that the Bogoliubov limit corresponds more closely to a physically meaningful definition of the vacuum persistence amplitude. In the meantime, the very existence of these ambiguities signals potential limitations of the in–out formalism in genuinely non–stationary backgrounds. As emphasized by Akhmedov et al. \cite{Akhmedov_2019,Akhmedov:2024qvi}, the more appropriate framework in such situations is the Keldysh–Schwinger technique, which provides access to well–defined observables such as the expectation value of the energy–momentum tensor rather than ill–defined particle numbers \cite{AKHMEDOV_2014LN}. Despite the fact that the Keldysh–Schwinger technique concerns a different class of physical observables, the phenomenon we report here is in fact generic in quantum field theory in de Sitter space: It is a manifestation of the well-known non-commutativity of the UV and IR limits \cite{Polyakov:2012uc}, which arises independently of the particular computational framework. When handling such divergent integrals, it is therefore important to rely on a physically motivated interpretation to select the appropriate order of limits or regularization schemes. Our unified analysis thus provides a complementary perspective, clarifying a conceptual issue that lies outside the natural scope of the Keldysh–Schwinger framework.

Finally, our results give rise to a broader implication: similar ambiguities are expected to arise whenever one attempts to extract vacuum instability from divergent integrals in dynamical backgrounds. The lesson is that particle production in curved spacetimes cannot be unambiguously defined without specifying a precise regularization prescription, and different choices may correspond to distinct physical setups. Therefore, predictions of vacuum instability in dynamical spacetimes should be treated with caution, and further study is needed to disentangle genuine physical effects from artifacts of regularization.

\section*{Acknowledgements}
This work was partially supported by the National Natural Science Foundation of China
(Grants No.12175008).

\bibliographystyle{unsrtnat}
\bibliography{ref}

\newpage
\appendix

\section{Detailed Derivation of the $\eta–k$ Integral from Eq.~(\ref{ImWint}) to Eq.~(\ref{ImWdS})}\label{AppendixA}

In this appendix, we present the detailed evaluation of the momentum and time integrations that appear from Eq.~(\ref{ImWint}) to Eq.~(\ref{ImWdS}).

To proceed, we isolate the part of the integral containing the $k$- and $\eta$-dependence,
\begin{align}
I=\int_{-\infty}^{\tau}\dd\eta\int_0^{\Lambda}\dd k \frac{k^{d-1}}{(-\eta)}\left(J_{i \mu }(-k \eta )^2-J_{-i \mu }(-k \eta )^2\right),
\end{align}
and change variables by defining $k=\frac{x}{(-\eta)}$. Then the integral becomes
\begin{align}
I=\int_{-\infty}^{\tau}\dd\eta\frac{1}{(-\eta)^{d+1}}\int_0^{-\eta\Lambda}\dd x x^{d-1}\left(J_{i \mu }(x )^2-J_{-i \mu }(x )^2\right).
\end{align}
By interchanging the order of integration, the integration domain splits into two regions:\\
(i) $0<x<\lambda\equiv -\tau\Lambda$, where the upper bound in $x$ is set by $-\eta\Lambda$;\\
(ii) $x>\lambda$, where the $\eta$-integration saturates at its upper limit $\eta=\tau$.
\begin{align}
I=&\int_0^{+\infty}\dd x x^{d-1}\left(J_{i \mu }(x )^2-J_{-i \mu }(x )^2\right) \int_{-\infty}^{\text{min}\{\tau,-x/\Lambda\}}\frac{\dd\eta}{(-\eta)^{d+1}} \\
=&\frac{(-\tau)^{-d}}{d}\int_0^{-\tau\Lambda}\dd x x^{d-1}\left(J_{i \mu }(x )^2-J_{-i \mu }(x )^2\right)+\frac{\Lambda^d}{d}\int_{-\tau\Lambda}^{+\infty}\dd x x^{-1}\left(J_{i \mu }(x )^2-J_{-i \mu }(x )^2\right).
\end{align}
This yields the expression
\begin{align}\label{Iint}
I=\frac{(-\tau)^{-d}}{d}\int_0^{\lambda}\dd x x^{d-1}\left(J_{i \mu }(x )^2-J_{-i \mu }(x )^2\right)+\frac{\Lambda^d}{d}\int_{\lambda}^{+\infty}\dd x x^{-1}\left(J_{i \mu }(x )^2-J_{-i \mu }(x )^2\right).
\end{align}
The remaining $x$-integrals are incomplete Weber–Schafheitlin–type integrals.
We evaluated these expressions using the integral formulas from the
{\href{https://functions.wolfram.com/Bessel-TypeFunctions/BesselJ/21/01/04/01/01/01/}{Wolfram Functions Site}}
and the corresponding
{\href{https://functions.wolfram.com/PDF/BesselJ.pdf}{Mathematica document}} 
\cite{WolframBesselJ}.
\begin{align}\label{Gint}
\int_0^{\lambda}\dd x x^{d-1}J_{-i\mu}(x)^2=\frac{\lambda^d(\lambda/2)^{-2i\mu}}{(d-2i\mu)\Gamma(1-i\mu)^2}\, _2F_3\left(\frac{1}{2}-i \mu ,\frac{d}{2}-i \mu ;1-i \mu ,\frac{d}{2}-i \mu +1,1-2 i \mu ;-\lambda ^2\right)
\end{align}
and
\begin{align}\label{Bint}
\int_{\lambda}^{+\infty}\dd x x^{-1}J_{-i\mu}(x)^2=\frac{i}{2\mu}-\frac{i(\lambda/2)^{-2i\mu}}{2\mu\Gamma(1-i\mu)^2}\, _2F_3\left(\frac{1}{2}-i \mu ,-i \mu ;1-i \mu ,1-i \mu ,1-2 i \mu ;-\lambda ^2\right)
\end{align}
together with their complex-conjugate counterparts obtained by $\mu\to -\mu$. The function ${}_2F_3$ appearing in (\ref{Gint}) and (\ref{Bint}) are defined in the main text (see Eqs. (\ref{GHF1}) and (\ref{GHF2})), as $\mathcal{F}_1(\mu,d,\lambda)$ and $\mathcal{F}_2(\mu,\lambda)$, respectively, for conciseness. Substituting these results back into (\ref{Iint}) yields the final closed-form expression quoted in (\ref{ImWdS})
\begin{align}
I=-i\left(\frac{1}{\mu}+\mathcal{F}(\mu,d,\lambda)+\tilde{\mathcal{F}}(\mu,d,\lambda)\right)
\end{align}
with
\begin{align}
\mathcal{F}(\mu,d,\lambda)=\frac{(\lambda/2)^{-2i\mu}}{\Gamma(1-i\mu)^2} \left(\frac{\mathcal{F}_1(\mu,d,\lambda)}{2\mu+id} -\frac{\mathcal{F}_2(\mu,\lambda)}{2\mu}\right)
\end{align}
defined in (\ref{FforGHFs}).

It is worth noting that Eq.(\ref{Iint}) separates the result into two manifestly distinct contributions: one proportional to the time cutoff $(-\tau)^{-d}$ and the other proportional to the momentum cutoff $\Lambda^{d}$. This structure provides an intuitive characterization of the integral's behavior in different limiting regimes of the trade-off parameter $\lambda$. Roughly speaking, when $\lambda \to 0^{+}$, the first integral over $x\in [0,\lambda]$ vanishes, and the contribution proportional to the momentum cutoff $\Lambda^{d}$ dominates the final result, which is characteristic of the result obtained from the Bogoliubov method.  Conversely, when $\lambda \to +\infty$, the second integral over $x\in [\lambda,\infty)$ is suppressed, and the dominant contribution arises from the part proportional to the time cutoff $(-\tau)^{-d}$, which is characteristic of the result obtained from the Green's function method. The precise analysis is provided in the main text.

\end{document}